\documentclass[
aps,prl
 amsmath,amssymb,
 reprint,%
]{revtex4-1}
\usepackage{chngcntr} 
\usepackage{graphicx}
\usepackage{float} 
\usepackage[font=small]{caption,subcaption} 
\usepackage{makecell} 
\usepackage{tabularx}
\usepackage{dcolumn}
\usepackage{bm}
\usepackage{tikz}
\usepackage[mathlines]{lineno}
\usepackage{mathptmx}
\usepackage{color}
\usepackage[colorlinks,linkcolor=blue]{hyperref}
\usepackage{physics}

\newcolumntype{Y}{>{\centering\arraybackslash}X}

\begin{document}
\title{EFIT-mini: An Embedded, Multi-task Neural Network-driven Equilibrium Inversion Algorithm}
\author{G.H. Zheng$^{1,4}$, S.F. Liu$^{1*}$, H.S. Xie$^{2}$, H.Y. Zhao$^{2}$, Y.P. Zhang$^{2}$, X. Gu$^{2}$, Z.Y. Chen$^{2}$, \\ T.T. Sun$^{2}$, Y.N. Xu$^{2}$, J. Li$^{2}$, D. Guo$^{2}$, R.Y. Tao$^{2}$, Y.J. Hu$^{3}$ and Z.Y. Yang$^{4}$ \\
\small{$^1$ School of Physics, Nankai University, Tianjin 300071, People's Republic of China}\\
\small{$^2$ ENN Science and Technology Development Company, Ltd., Langfang 065001, People's Republic of China}\\
\small{$^3$ Institute of Plasma Physics, Chinese Academy of Sciences, Hefei 230031, People's Republic of China}\\
\small{$^4$ Southwestern Institute of Physics, Chengdu 610041, People's Republic of China}\\
\small{$^*$Email : lsfnku@nankai.edu.cn}\\}

\begin{abstract}
Equilibrium reconstruction, which infers internal magnetic fields, plasmas current, and pressure distributions in tokamaks using diagnostic and coil current data, is crucial for controlled magnetic confinement nuclear fusion research. However, traditional numerical methods often fall short of real-time control needs due to time-consuming computations or iteration convergence issues. This paper introduces EFIT-mini, a novel algorithm blending machine learning with numerical simulation. It employs a multi-task neural network to replace complex steps in numerical equilibrium inversion, such as magnetic surface boundary identification, combining the strengths of both approaches while mitigating their individual drawbacks. The neural network processes coil currents and magnetic measurements to directly output plasmas parameters, including polynomial coefficients for $p'$ and $ff'$, providing high-precision initial values for subsequent Picard iterations. Compared to existing AI-driven methods, EFIT-mini incorporates more physical priors (e.g., least squares constraints) to enhance inversion accuracy. Validated on EXL-50U tokamak discharge data, EFIT-mini achieves over 98\% overlap in the last closed flux surface area with traditional methods. While traditional methods take seconds per time slice for $129 \times 129$ resolution reconstructions, EFIT-mini's neural network and full algorithm compute single time slices in just 0.11ms and 0.36ms, respectively, representing a three-order-of-magnitude speedup. This innovative approach leverages machine learning's speed and numerical algorithms' explainability, offering a robust solution for real-time plasmas shape control and potential extension to kinetic equilibrium reconstruction. Its efficiency and versatility position EFIT-mini as a promising tool for tokamak real-time monitoring and control, as well as for providing key inputs to other real-time inversion algorithms.
\end{abstract}

\maketitle

\section{Introduction}

Magnetic equilibrium reconstruction is a method that uses magnetic diagnostic signals from tokamak discharge experiments as constraints, combined with data on coil currents during the discharge, to compute the spatial distribution of current density, magnetic field, and flux surfaces in equilibrium, as well as the profiles of current and pressure. Inferring the internal electromagnetic state and distribution solely from external electromagnetic information is an "inverse problem". This requires preparing a suitable flux surface distribution as the trial solution for iterative refinement until a balanced solution consistent with both external constraints and internal distribution is achieved. Currently, the popular equilibrium reconstruction algorithm is EFIT \cite{lao1985reconstruction}, which solves the GS equation \cite{grad1958hydromagnetic, shafranov1966plasma} using Picard iteration to update flux surfaces. It has been implemented on numerous devices, including DIII-D, JET, NSTX, EAST, KSTAR, START, C-MOD, TORE SUPRA, HL-2A, HL-3, QUEST, MAST, and EXL-50U \cite{lao2005mhd, o1992equilibrium, sabbagh2001equilibrium, jinping2009equilibrium, li2013kinetic, park2011kstar, jiang2021kinetic, appel2001equilibrium, appel2006unified, in2000resistive, zwingmann2003equilibrium, li2011efit, hongda2006study, xue2019equilibrium, berkery2021kinetic}, making significant contributions to tokamak discharge experiments.
While magnetic equilibrium reconstruction, especially in real-time, is crucial for tokamak discharge experiments, existing algorithms have yet to fully meet experimental needs, which mainly fall into two categories:

$\bullet$ Real-time performance: The EFIT algorithm takes seconds to reconstruct flux surfaces at 129 $\times$ 129 resolution for a single time slice. For real-time control of large tokamaks, this needs to be optimized to $\leq$10ms per slice, and even $\leq$1ms for smaller devices.

$\bullet$ Stability: Iterative solving can lead to instabilities from physical factors (e.g., vertical-displacement-effect-caused magnetic axis shifts) and numerical factors (e.g., least-squares rounding errors, Picard iteration initial solution selection). In ITER, which will operate at up to 15 MA plasmas current and 500 MW fusion power, real-time magnetic control failure could have irreversible consequences.

To boost equilibrium reconstruction speed, prior work has optimized the reconstruction code. Early efforts include DIII-D's rt-EFIT \cite{ferron1998real}, which assumes slow flux surface variation, uses previous time slice's flux distribution as the trial at this slice, reduces iterations to once per slice, and skips boundary flux calculation, achieving $<$1ms reconstruction at 65 $\times$ 65 resolution. Similar work also has been conducted on HL-2A \cite{rui2018acceleration}. Recently, EAST's P-EFIT leveraged GPU parallel computing, optimizing boundary flux search and Picard iteration for divertor configurations, reducing 129 $\times$ 129 resolution reconstruction time to about 0.5ms \cite{huang2020gpu}. However, the slow variation assumption introduces stability issues, as instabilities in one slice can propagate to subsequent ones. Additionally, these methods either avoid or simplify boundary flux surface identification, a challenging task for traditional algorithms, especially for limiter-divertor transitions or advanced divertor configurations.

Machine learning offers a viable path to enhancing both reconstruction speed and stability. Models trained with added noise during training tend to be more robust. Early applications of machine learning to equilibrium reconstruction include KSTAR's work using coordinates $(R, Z)$ as input to predict flux values at the point \cite{joung2019deep}, and DIII-D's approach with convolutional layers to directly output flux distributions, better preserving spatial physical information \cite{lao2022application}. DIII-D's EFIT-Prime \cite{10.1063/5.0213625, 10.1063/5.0213609} later reduced network size by mimicking EFIT's current density construction, improving speed significantly. KSTAR's GS-DeepNet \cite{joung2023gs} used physics-informed neural networks (PINNs) for unsupervised learning, incorporating physical information without generating EFIT data. Similar work has been done on Globus-M2, NSTX-U, EAST, HL-3 and simulations for ITER-like devices \cite{mitrishkin2021new, wai2022neural, lu2023fast, zheng2024real, bonotto2024reconstruction}. However, most existing work focuses on either adding physical information to loss functions for accuracy or simplifying models for speed, neglecting the balance between them. Moreover, these models are end-to-end, taking raw magnetic measurements as input and outputting final results like flux surfaces and profiles, with no intervention possible during inference. While this approach doesn't require deep EFIT knowledge, it will not only limit accuracy enhancement to intermediate parameter constraints by loss functions but also inevitably increase network size and reduce speed when outputting entire flux distributions.

To harness the high stability of neural networks, control network size for speed, and leverage equilibrium reconstruction principles, a real-time equilibrium reconstruction algorithm embedding the neural network, named EFIT-mini has been developed in this paper. It combines the strengths of numerical and machine learning methods while mitigating their weaknesses.

Unlike other neural network-based approaches, EFIT-mini uses neural networks only for the most challenging numerical steps, allowing better use of intermediate parameters' physical constraints for higher accuracy. Instead of directly outputting high-resolution flux surfaces, the neural network in EFIT-mini outputs a few key parameters, reducing complexity and boosting speed. The algorithm can compute flux surfaces at any resolution determined by the matrix dimension in Picard iteration for solving the GS equation. This also allows using offline EFIT results at any resolution for training. Besides, retaining the Picard iteration process helps correct minor errors in the output flux geometry parameters.

Compared to numerical algorithms, EFIT-mini outputs several plasmas parameters and least-squares solutions directly via neural networks from the current slice, rather than relying on the previous slice's flux solution. Combined with Gaussian noise added during training, EFIT-mini reduces instability accumulation and rounding errors and enhancing robustness. For 129 $\times$ 129 resolution flux surface reconstruction, the neural network and the entire EFIT-mini process take about only 0.11ms and 0.36ms per time slice, respectively, meeting real-time requirements.

The following sections are structured as follows: Section \ref{sec:methods} analyzes the offline EFIT algorithm's basic steps to identify time-consuming or iteration-unstable steps for potential replacement with machine learning methods. It also introduces EFIT-mini's structure and details on adding physical information to improve accuracy, including multi-task learning and physical loss strategies. Section \ref{sec:results} presents the neural network's performance and the equilibrium reconstruction algorithm's accuracy and speed. Finally, Section \ref{sec:discussion} summarizes the work and discusses potential improvements and optimizations for EFIT-mini.

\section{Methods}
\label{sec:methods}

This section outlines the implementation of EFIT-mini. We start by analyzing the offline EFIT code to identify performance and accuracy bottlenecks, then determine suitable replacements with neural networks. Following this, we detail the neural network implementation and training, emphasizing physical information integration to enhance model accuracy.

\subsection{Analysis on offline-EFIT code}
\begin{figure*}
    \centering
    \includegraphics[width=.45\textwidth]{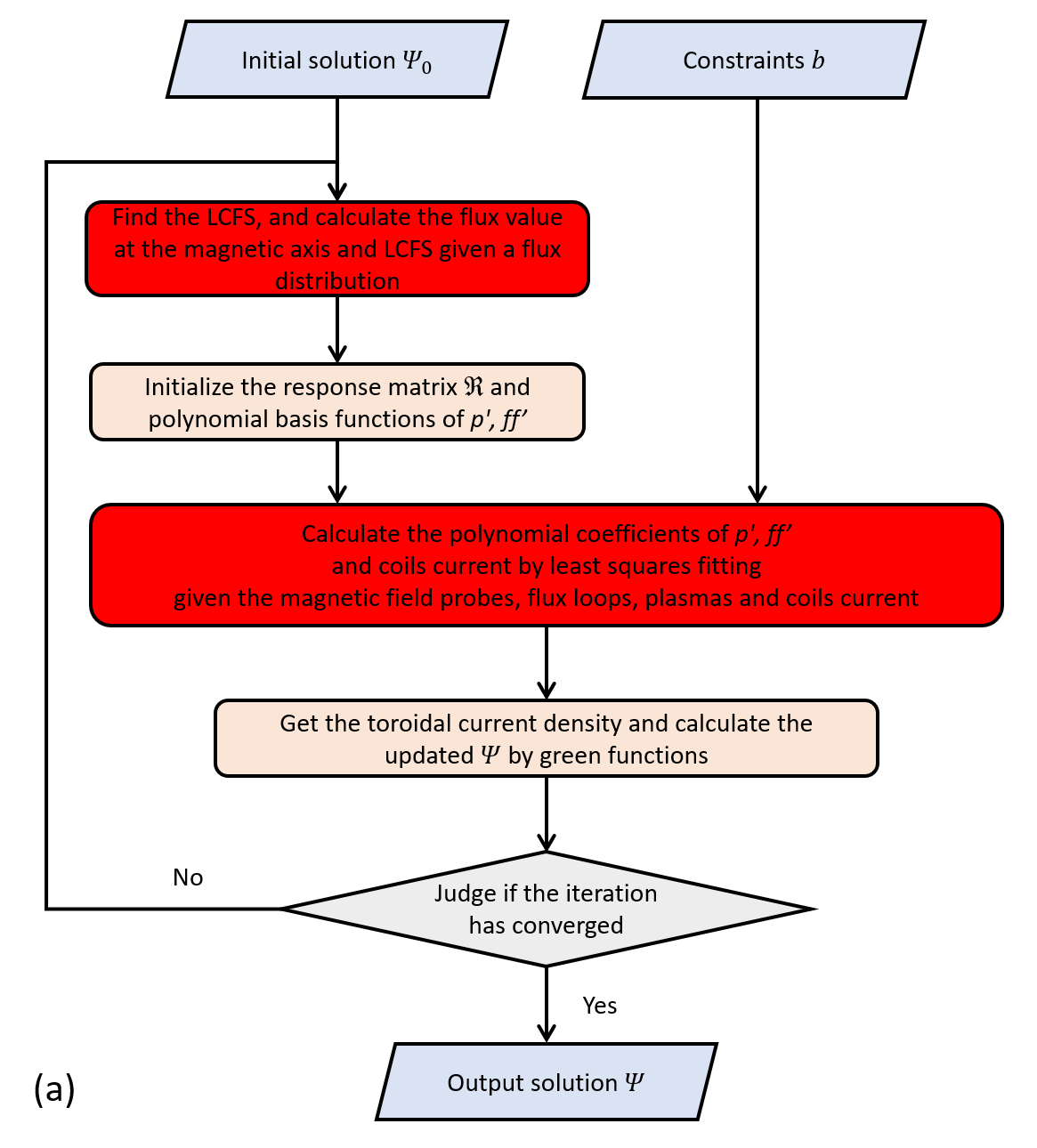}
    \includegraphics[width=.45\textwidth]{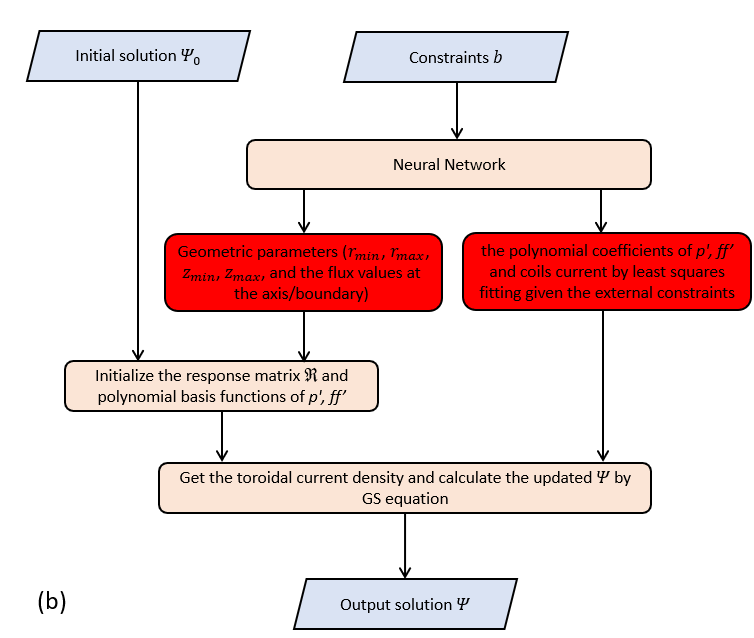}
    \caption{(a) Flowchart of the equilibrium reconstruction algorithm EFIT. (b) Flowchart of the EFIT-mini equilibrium inversion.}
    \label{fig:principle}
\end{figure*}

In this subsection, we analyze the offline EFIT algorithm and identify challenging steps for real-time applications. Figure \ref{fig:principle}(a) illustrates the offline EFIT process.

Here, \( p \) and \( f \) represent the plasmas pressure profile and poloidal current profile. The algorithm uses Picard iterations to find a convergent solution for the poloidal flux distribution. For a trial solution \( \Psi_0 \) and the \( i \)-th iteration solution \( \Psi_i \), the algorithm locates the magnetic axis and the last closed flux surface (LCFS) to normalize the flux \( \tilde{\Psi} \) and construct the current density basis functions. In EFIT, \( p' \) and \( ff' \) are modeled as polynomials of \( \tilde{\Psi} \). The current density, derived from these polynomials, is used to update the flux distribution via Green's functions or the GS equation, completing one Picard iteration. This process repeats until convergence.

For real-time control, equilibrium reconstruction must occur every millisecond. Under the assumption of gradual flux variation, the previous iteration's solution can seed the next, reducing iterations to once and significantly speeding up the algorithm. However, three main challenges remain:

$\bullet$ Rapidly determining the boundary flux, magnetic axis flux, and LCFS shape for a given flux solution. Existing solutions, like P-EFIT, still struggle with limiter-divertor transitions and complex divertor configurations.

$\bullet$ Robustly determining current density polynomial coefficients and coil currents under given \( p', ff' \) distributions and external constraints. This involves solving an over-determined equation \( \mathfrak{R}X = b \) via least squares, which can be unstable and hard to parallelize.

$\bullet$ Efficiently inverting the flux distribution from the current density via Picard iteration, which involves large matrix operations incompatible with real-time speeds.

We propose replacing the first two challenges with neural networks, leveraging their stability and parallel computing capabilities. The third challenge remains handled by numerical methods, due to the fact that neural networks are also implemented in principle of large matrix operations, and therefore they will certainly not perform better than numerical methods in this aspect. Besides, there has been some related work on the simplicity of matrix computations of the inversion in just about 0.1ms per slice at 129 $\times$ 129 resolution\cite{huang2017fast}.

Figure \ref{fig:principle}(b) presents the EFIT-mini inversion algorithm. Unlike numerical methods, EFIT-mini uses neural networks to directly output least squares solutions and plasmas parameters, which enhances the stability greatly.

\subsection{Data}
The model's inputs and outputs are shown in table \ref{tab:io}. Inputs include flux loops, magnetic probes, and currents from coils and plasmas, corresponding to the right-hand side of the least squares equation in numerical algorithms. Outputs include the least squares solution and plasmas geometric parameters.

\begin{table}[h]
    \centering
    \caption{Model inputs and outputs.}
    \begin{tabular}{lll}
        \hline
        Name & Type & Num. of channels \\
        \hline
        Constraints & Input & 161 \\
        \hline
        Flux Loops & Input & 47 \\
        Magnetic probes & Input & 100 \\
        Current of PF coils & Input & 12 \\
        Current of CS coil & Input & 1 \\
        Current of plasmas & Input & 1 \\
        \hline
        Solutions of LSQ & Output1 & 15 \\
        \hline
        Coefficients of \( p' \) & Output1 & 1 \\
        Coefficients of \( ff' \) & Output1 & 1 \\
        Current of PF coils & Output1 & 12 \\
        Current of CS coil & Output1 & 1 \\
        \hline
        Geometric parameters & Output2 & 8 \\
        \hline
        \( \Psi \) on LCFS & Output2 & 1 \\
        \( \Psi \) at magnetic axis & Output2 & 1 \\
        Upper triangularity & Output2 & 1 \\
        Lower triangularity & Output2 & 1 \\
        Maximum \( r \) of the LCFS & Output2 & 1 \\
        Minimum \( r \) of the LCFS & Output2 & 1 \\
        Maximum \( z \) of the LCFS & Output2 & 1 \\
        Minimum \( z \) of the LCFS & Output2 & 1 \\
        \hline
    \end{tabular}
    \label{tab:io}
\end{table}

All training, validation, and test datasets originate from offline inversion results of the EXL-50U tokamak real discharge experiments, comprising 355 shots and 206 543 slices from shot \#6404 to \#7512. Data is split into training (70\%), validation (10\%), and test (20\%) sets by shot number. Preprocessing steps include:

$\bullet$ Filtering data to exclude low plasmas current phases (<80 kA), which always include disruptions or rapidly changed plasmas currents with significant wall eddy currents.

$\bullet$ Adding Gaussian noise to inputs to improve model robustness and simulate diagnostic errors, with noise magnitude based on probe calibration errors.

$\bullet$ Z-score normalization of all data channels.

\subsection{Structure and Metrics of NN in EFIT-mini}
\label{subsec:metrics}
\begin{figure}[H]
    \centering
    \includegraphics[width=.45\textwidth]{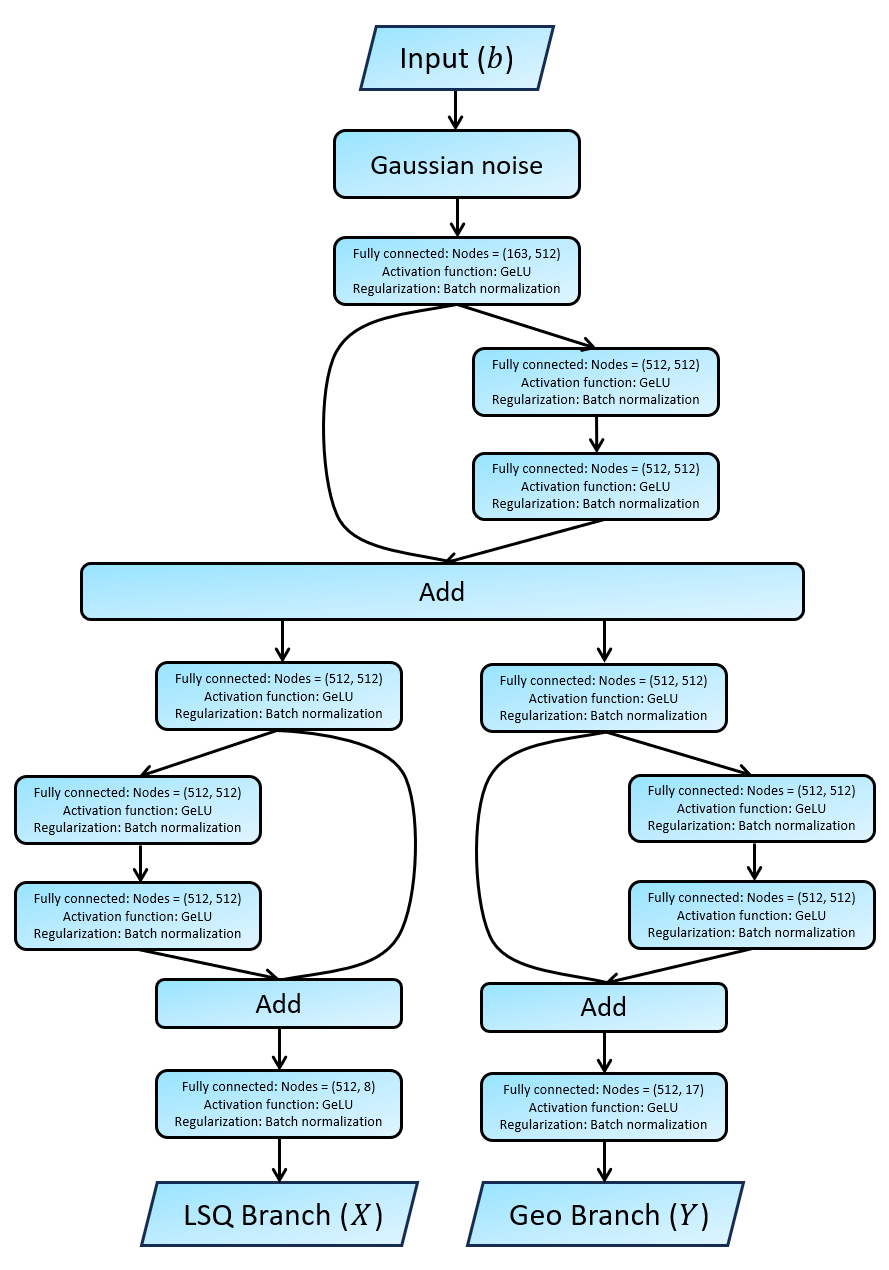}
    \caption{Neural network structure in EFIT-mini. Here, \( b \) is the input (external magnetic measurements), and \( X \), \( Y \) are the outputs for least squares and geometric parameters.}
    \label{fig:mini_structure}
\end{figure}

The neural network structure in EFIT-mini is depicted in figure \ref{fig:mini_structure}.

The network employs a multi-task learning structure, which has been proved effective on sharing informations between branches, enhancing robustness and output accuracy \cite{zheng2024real}. As an embedded module within the numerical algorithm, it effectively incorporates physical constraints from inversion parameters. More detailed, this network can introduce at least four key physical constraints:

(1) Least squares constraint
\begin{equation}
    ||\mathfrak{R}(\tilde\Psi, \tilde Y)X - b|| + ||\mathfrak{R}(\tilde\Psi, Y)\tilde X - b|| + ||\mathfrak{R}(\tilde\Psi, Y)X - b||
    \label{eq:lsq_constraint}
\end{equation}

(2) Least squares PINN constraint
\begin{equation}\begin{aligned}
    || & \mathfrak{R}(\tilde\Psi, \tilde Y) \dfrac{\partial X}{\partial b} + X \dfrac{\partial\mathfrak{R}(\tilde\Psi, \tilde Y)}{\partial b} - E|| + \\
    || & \mathfrak{R}(\tilde\Psi, Y) \dfrac{\partial X}{\partial b} + X \dfrac{\partial\mathfrak{R}(\tilde\Psi, Y)}{\partial b} - E||
\end{aligned}\end{equation}

(3) GS equation constraint
\begin{equation}\begin{aligned}
    || & \Delta^*\tilde\Psi + \mu_0RJ_\varphi(\tilde\Psi, \tilde X, Y)|| + \\
    || & \Delta^*\tilde\Psi + \mu_0RJ_\varphi(\tilde\Psi, X, \tilde Y)|| + \\
    || & \Delta^*\tilde\Psi + \mu_0RJ_\varphi(\tilde\Psi, X, Y)||
    \label{eq:gs_constraint}
\end{aligned}\end{equation}

(4) Green's function constraint
\begin{equation}\begin{aligned}
    || & \int G_JJ_\varphi(\tilde\Psi, \tilde X, Y)\text dA + G_CI_C(\tilde X) - \tilde \Psi|| + \\
    || & \int G_JJ_\varphi(\tilde\Psi, X, \tilde Y)\text dA + G_CI_C(X) - \tilde \Psi|| + \\
    || & \int G_JJ_\varphi(\tilde\Psi, X, Y)\text dA + G_CI_C(X)- \tilde \Psi||
\end{aligned}\end{equation}

Here, \( E \) is the identity matrix, \( G_J \) and \( G_C \) are Green's function matrices for plasmas current density \( J_\varphi \) and coil current \( I_C \), and \( dA \) is the area element. \( X, Y \) are network predictions, while \( \tilde X, \tilde Y \) are offline EFIT labels. Notably, the network does not directly output \( \Psi \), so \( \tilde\Psi \) from EFIT is used in constraints. In fact, it is hopeful to add a \( \Psi \) output branch and enable unsupervised learning, but it would complicate the network significantly.

Each loss type $\mathfrak{L}$ (except PINN) comprises three terms because they cover combinations of both network branches. Three terms can be expressed as $||\mathfrak{L}(X, \tilde Y) - \mathfrak{L}(\tilde X, \tilde Y)||$, $||\mathfrak{L}(\tilde X, Y) - \mathfrak{L}(\tilde X, \tilde Y)||$, and $||\mathfrak{L}(X, Y) - \mathfrak{L}(\tilde X, \tilde Y)||$, respectively. The former two terms behave like the data loss to enhance their own branch's accuracy, while the last one make informations flow between branches. This design stabilizes training and leverages multi-task learning. However, excessive constraints risk local minima and increased training complexity. Given the offline EFIT data generation using GS-equation-based Picard iterations, we selectively use least squares and GS equation constraints, as shown in table \ref{tab:loss_cfg}, balancing training duration and physical relevance.

\begin{table}[h]
    \centering
    \caption{Loss functions, weights, and steps used in model training.}
    \begin{tabular}{lll}
        \hline
        Constraint & Weight & Step \\
        \hline
        $||X - \tilde X||$ & 1 & 0 \\
        $||Y - \tilde Y||$ & 1 & 0 \\
        $||\mathfrak{R}(\tilde\Psi, \tilde Y)X - b||$ & 1000 & 12500 \\
        $||\mathfrak{R}(\tilde\Psi, Y)\tilde X - b||$ & 650 & 12500 \\
        $||\mathfrak{R}(\tilde\Psi, Y)X - b||$ & 650 & 12500 \\
        $||\Delta^*\tilde\Psi + \mu_0RJ_\varphi(\tilde\Psi, X, \tilde Y)||$ & 52 & 12500 \\
        $||\Delta^*\tilde\Psi + \mu_0RJ_\varphi(\tilde\Psi, X, Y)||$ & 80 & 12500 \\
        \hline
    \end{tabular}
    \label{tab:loss_cfg}
\end{table}

Table \ref{tab:training_cfg} lists other key hyperparameters used in training, including a learning rate that decreases with training steps.

\begin{table}[h]
    \centering
    \caption{Key hyperparameters used in model training.}
    \begin{tabular}{lll}
        \hline
        Name & Value \\
        \hline
        Learning rate & $10^{-3} \times 0.5^{\text{step} / 10^4}$ \\
        Nodes for each FC layer & 512 \\
        Activation function & GeLU \\
        Batch size & 128 \\
        Optimizer & AdamW \\
        \hline
    \end{tabular}
    \label{tab:training_cfg}
\end{table}

\section{Results}
\label{sec:results}

\begin{figure}[h]
    \centering
    \includegraphics[width=.45\textwidth]{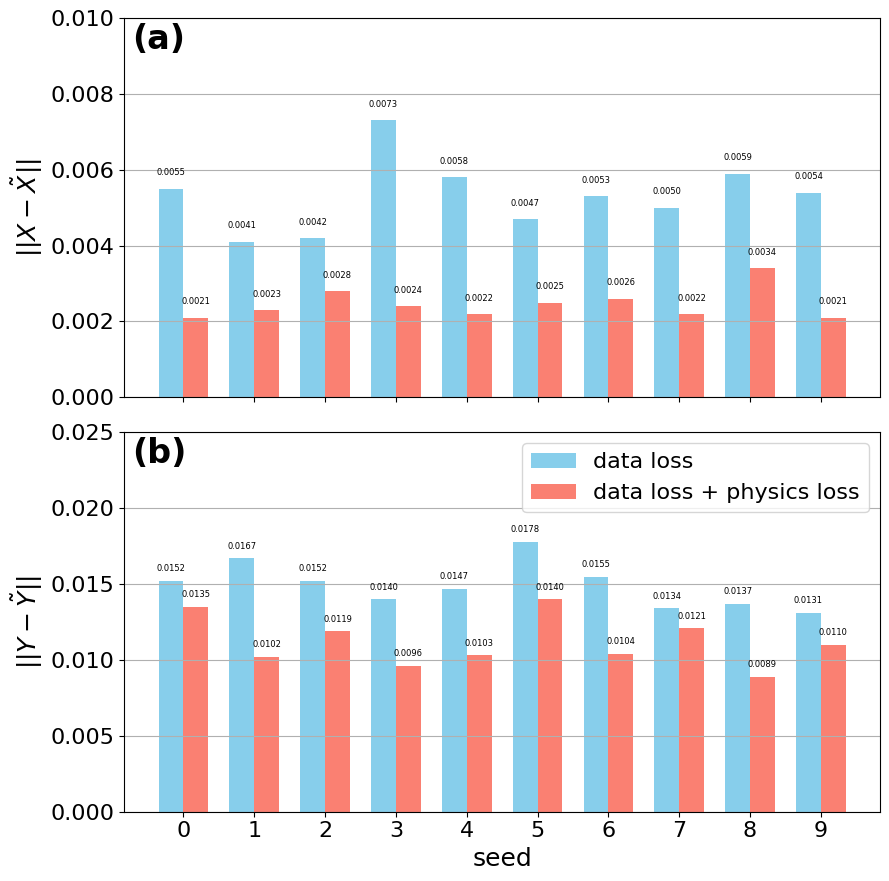}
    \caption{Comparison of training effects using pure data loss (blue) and combined data and physical loss (red) in ten experiments, assessed by the RMSE between model predictions and true values of $X$ and $Y$.}
    \label{fig:data_vs_extra}
\end{figure}

This section presents the performance, computational speed, and comparative evaluation of EFIT-mini and its embedded neural network against the "offline EFIT" results in various aspects. We should emphasize that the "offline EFIT" mentioned in this section is a python-port of the equilibrium inversion algorithm \cite{heq_github}, not the original EFIT code.  By employing a multi-task learning architecture and incorporating physical losses as outlined in section \ref{subsec:metrics}, the precision of the neural network has been significantly enhanced. To demonstrate this, we have trained the model using ten different random seeds, which influence the initial weight distributions and the sequences of training samples presented to the model. For each seed, we conducted two training sessions: one using pure data loss and another combining data loss with physical loss. The final training outcomes are depicted in figure \ref{fig:data_vs_extra}.

The results indicate that training with the combination of data loss and physical constraints yields superior outcomes compared to using pure data loss. Notably, the optimization effect on $X$ is more pronounced, with an average loss reduction of 52.5\%, while $Y$'s loss decreases by an average of 24.8\%.

\begin{figure}[H]
    \centering
    \includegraphics[width=.45\textwidth]{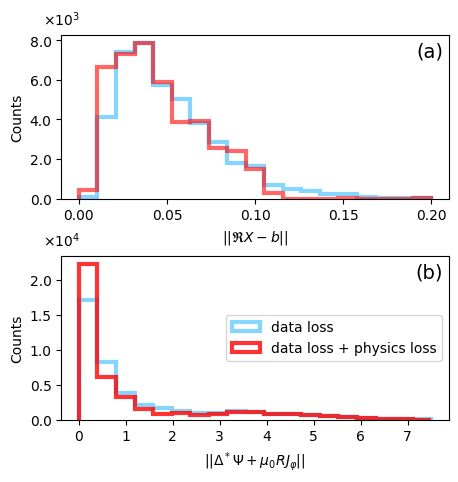}
    \caption{Distribution of (a) least-squares residuals and (b) GS equation residuals on the test set for models trained with data loss and physical loss.}
    \label{fig:gs_and_chi}
\end{figure}

Furthermore, figure \ref{fig:gs_and_chi} illustrates the distribution of the least-squares residuals $||\mathfrak{R}(\tilde\Psi, Y)X - b||$ and GS equation residuals $||\Delta^*\tilde\Psi + \mu_0RJ_\varphi(X, Y)||$ for model outputs $X, Y$ on the test set, using both pure data loss and added physical loss training methods. The comparison reveals that the model with physical loss constraints exhibits obviously smaller residuals for both metrics, indicating that physical constraints enhance the optimization by guiding the neural network to update its internal weights in a more physically consistent manner.

\begin{figure*}
    \centering
    \includegraphics[width=.9\textwidth]{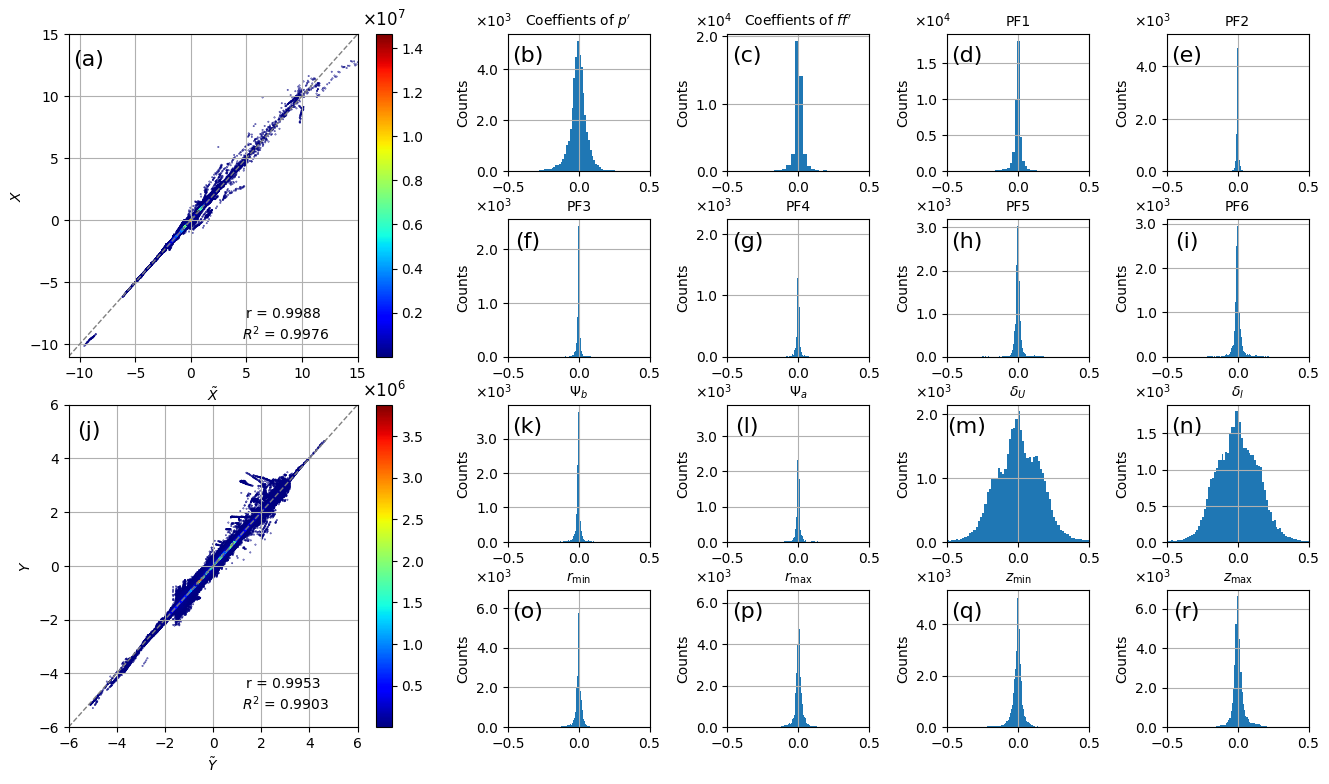}
    \caption{(a) Comparison of model output $X$ with offline EFIT, with the $y=x$ reference line shown in gray dashed line. (b)-(i) Relative errors between model predictions and reference values for selected parameters in $X$, including the polynomial coefficients of $p'$ and $ff'$ (1, 1-order in this experiment) and the inverted PF coil currents. (j) Comparison of model output $Y$ with offline EFIT. (k)-(r) Relative errors for all specific parameters in $Y$, including boundary flux $\Psi_b$, magnetic axis flux $\Psi_a$, upper and lower triangularities $\delta_U, \delta_L$, and the maximum and minimum values of $r$ and $z$ on the LCFS. Note: Results are based on normalized data for intuitive comparison of relative errors across different outputs.}
    \label{fig:mini_vs_efit}
\end{figure*}

Figure \ref{fig:mini_vs_efit} displays the correlation plots and relative error comparisons of the model's outputs $X$ and $Y$ with the corresponding results from offline EFIT. These results were obtained using the best-performing seed with both data and physical constraints. The network's outputs exhibit a high degree of consistency with offline EFIT, indicating the neural network's capability to address the challenging aspects of EFIT's numerical algorithm. It is also observed that the model's prediction accuracy for $X$ surpasses that for $Y$. This may be attributed to the fact that in $Y$, only the boundary and magnetic axis flux participate in the gradient propagation of all physical losses. The maximum and minimum $r, z$ of the LCFS are involved in partial gradient propagation when constructing the current density basis functions, while the triangularity parameters do not contribute to the physical loss calculations and thus are not further optimized by the added physical constraints. Figure \ref{fig:mini_vs_efit} (k)-(r) detail the relative errors of each specific output in $Y$ compared to offline EFIT, showing that the errors of $\Psi_b$ and $\Psi_a$ are notably smaller than other outputs, supporting the aforementioned conclusion. Nevertheless, the retention of the Picard iteration process in EFIT-mini allows for some correction of these deviations during the final magnetic surface inversion, even when the model's shape parameters are occasionally less precise.

\begin{figure}
    \centering
    \includegraphics[width=.45\textwidth]{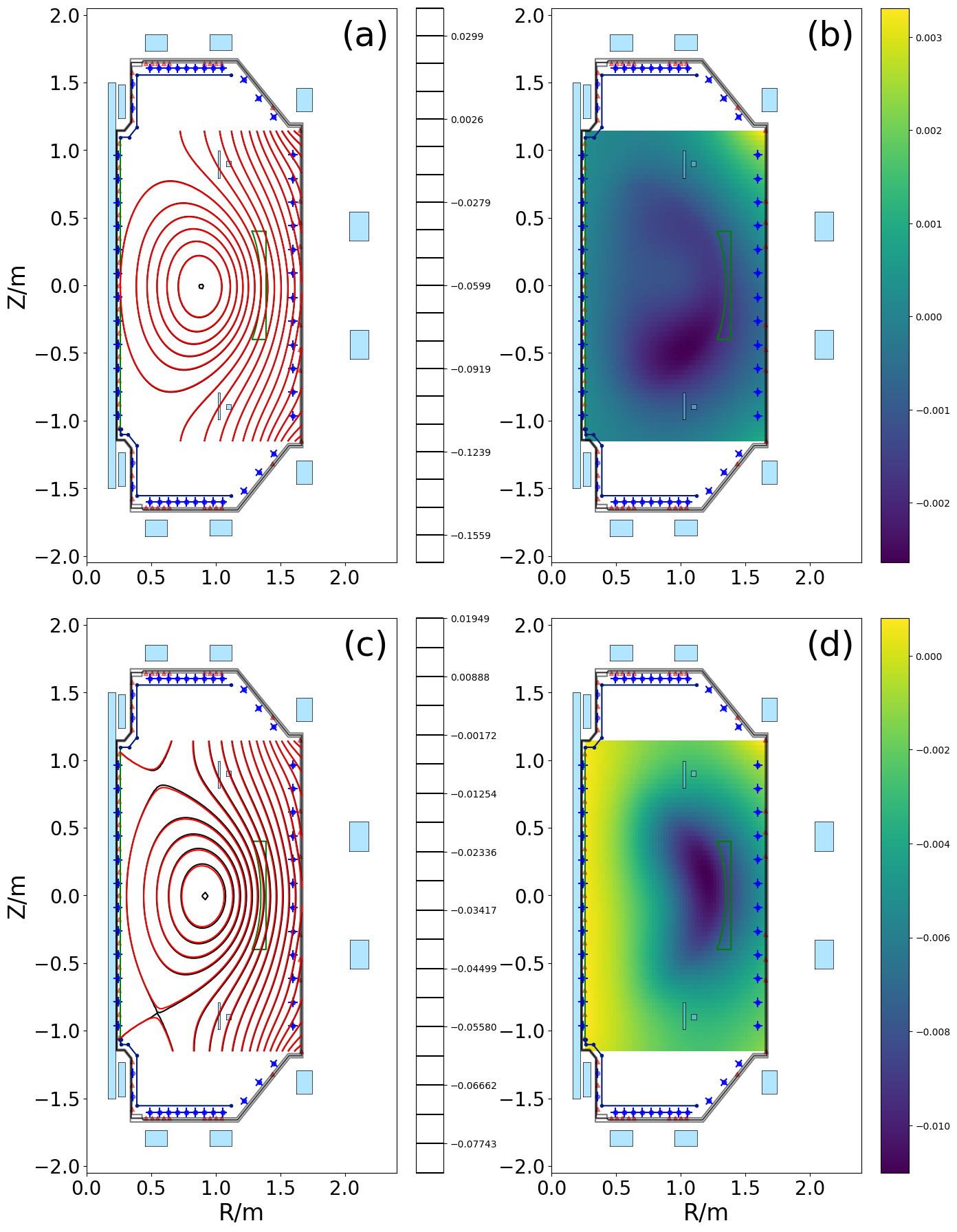}
    \caption{(a) Comparison of magnetic surface distributions for a typical limiter shot (\#6460 at 500ms). Black lines denote offline EFIT results, while red lines represent EFIT-mini results. (b) Residual comparison of magnetic surface distributions under the limiter configuration, defined as $|\Delta| = |\Psi_\text{nn} - \Psi_\text{EFIT}| / \max(|\Psi_\text{EFIT}|)$. (c)(d) Comparison of magnetic surface and residual distributions for a typical divertor shot (\#6404 at 800ms).}
    \label{fig:psi}
\end{figure}

In real-time inversion, the algorithm uses the magnetic surface from the previous time slice as an initial trial solution for the current slice, iterating once to invert the current magnetic surface. To validate the model's accuracy for this algorithm, we performed real-time inversion on complete discharges from the test set using EFIT-mini, selecting slices from a typical limiter shot (\#6460 at 500ms) and a typical divertor shot (\#6404 at 800ms) for illustration. The comparison of the magnetic surface distributions inverted by the algorithm with those from offline EFIT is shown in figure \ref{fig:psi}.

\begin{figure}
    \centering
    \includegraphics[width=.45\textwidth]{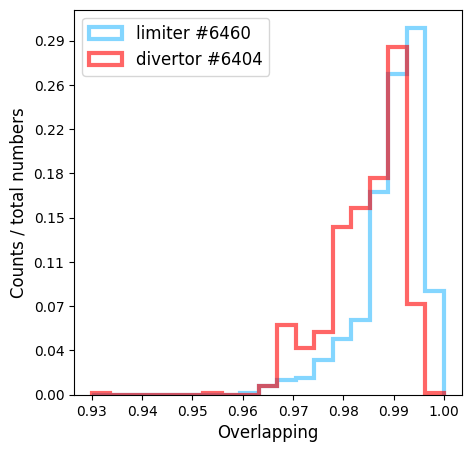}
    \caption{Distribution of overlap ratios for LCFS areas computed by EFIT-mini and offline EFIT in limiter and divertor configurations.}
    \label{fig:overlap}
\end{figure}

The results show an excellent agreement between the magnetic surfaces inverted by EFIT-mini and those from offline EFIT. To further highlight this consistency, we focus on the overlap of the LCFS from both methods. The overlap ratio $O$ is defined as:

\begin{equation}
    O = \dfrac{2S_O}{S_\text{EFIT} + S_\text{mini}},
\end{equation}

where $S_\text{EFIT}$ and $S_\text{mini}$ are the cross-sectional areas of the magnetic surfaces computed by offline EFIT and EFIT-mini, respectively, and $S_O$ is the overlapping area of their LCFS. The closer $O$ is to 1, the better performance EFIT-mini reaches. Figure \ref{fig:overlap} presents the distribution of overlap ratios of the LCFS computed by EFIT-mini and offline EFIT for the majority of time slices in shots \#6404 and \#6460 (excluding the first 20 slices of each shot to allow the real-time algorithm to converge). The average overlap ratios for limiter and divertor configurations are 0.990 and 0.985, respectively. The slightly lower overlap ratio for the divertor configuration is attributed to the smaller gradient of the magnetic flux function near the X-point, where minor numerical changes can lead to obvious shifts in the LCFS.

\begin{figure}[H]
    \centering
    \includegraphics[width=.45\textwidth]{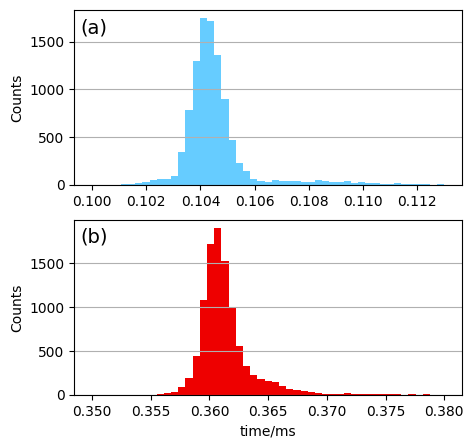}
    \caption{(a) Distribution of inference times per time slice for the embedded neural network in EFIT-mini. (b) Distribution of computation times per slice for the entire EFIT-mini algorithm.}
    \label{fig:compute_time}
\end{figure}

Finally, figure \ref{fig:compute_time} shows the distribution of the computation time per slice for both the model inference component and the entire EFIT-mini algorithm. The algorithm was implemented by NVIDIA (R) TensorRT framework to reduce the runtime as far as possible. Testing was conducted on NVIDIA (R) A100 Tensor Core GPU, with 10 000 time slices preprocessed before recording the time distribution for another 10 000 slices. The results indicate an average inversion time of approximately 0.36ms for the entire algorithm, with the neural network inference averaging less than 0.11ms and exhibiting stable computation times without timeouts. The computation speed of EFIT-mini is slightly faster than P-EFIT and 1 000 times faster than offline EFIT. Thus, EFIT-mini fully meets the requirements for real-time magnetic balance inversion in terms of both speed and stability.

\section{Discussion}
\label{sec:discussion}
We have developed EFIT-mini, a real-time inversion algorithm based on an embedded neural network. This network, trained on EXL-50U discharge data, outputs key parameters for magnetic surface reconstruction, enabling rapid magnetic equilibrium rebuilding via a single Picard iteration. The network's embedded and multi-task learning structure allows for enhanced precision through additional physical constraints. Moreover, by retaining the final Picard process, the algorithm reduces parameters, boosts speed, and mitigates errors from magnetic surface parameters. In essence, EFIT-mini combines the strengths of neural networks and numerical methods while avoiding their respective drawbacks as far as possible. The model maintains high consistency with offline EFIT results across different discharge configurations, achieving over 98\% coincidence in the LCFS. It also offers rapid inference times—approximately 0.11ms per time slice for the network and 0.36ms per time slice for the entire algorithm—meeting real-time requirements for precision and speed.

There is room for further optimization of the current model and real-time reconstruction algorithm. The model's least-squares parameters currently include polynomial coefficients for $p'$ and $ff'$ based on the normalized flux $\tilde\Psi$. In the future, the algorithm could be extended to output various profiles and incorporate current and kinetic reconstruction constraints.

Finally, our design philosophy is worth reiterating: we aim to use neural networks only for challenging steps in numerical algorithms. Compared to end-to-end surrogate models, this approach offers two key advantages: reduced model parameters and the ability to incorporate more physical informations, which can significantly enhance model accuracy, speed, and development efficiency.

\begin{acknowledgments}
This work is supported by National Natural Science Foundation of China under Grant No. 12275142, the National Key R\&D Program of China under Grant No. 2024YFE03020001, and ENN Science and Technology Development Company, Ltd.
\end{acknowledgments}

\bibliographystyle{unsrt}
\bibliography{ref}

\end{document}